# Asteroid spin-axis longitudes from the Lowell Observatory database

E. Bowell[1], D. A. Oszkiewicz[2,5], L. H. Wasserman[1], K. Muinonen[2,3], A. Penttilä[2], D. E. Trilling[4]

[1] Lowell Observatory, 1400 West Mars Hill Road, Flagstaff, AZ 86001, U.S.A.
[2] Department of Physics, P.O. Box 64, FI-00014 University of Helsinki, Finland.
[3] Finnish Geodetic Institute, P.O. Box 15, FI-02431 Masala, Finland.
[4] Department of Physics and Astronomy, Northern Arizona University,
P.O. Box 6010, Flagstaff, AZ 86011, U.S.A.
[5] Institute Astronomical Observatory, Faculty of Physics, Adam Mickiewicz University, Słoneczna 36, 60-286 Poznań, Poland

**Abstract**
By analyzing brightness variation with ecliptic longitude and using the Lowell Observatory photometric database, we estimate spin-axis longitudes for more than 350 000 asteroids. Hitherto, spin-axis longitude estimates have been made for fewer than 200 asteroids. We investigate longitude distributions in different dynamical groups and asteroid families. We show that asteroid spin-axis longitudes are not isotropically distributed as previously considered. We find that the spin-axis longitude distribution for main-belt asteroids is clearly non-random, with an excess of longitudes from the interval 30°-110° and a paucity between 120°-180°. The explanation of the non-isotropic distribution is unknown at this point. Further studies have to be conducted to determine if the shape of the distribution can be explained by observational bias, selection effects, a real physical process or other mechanism.

## 1 Introduction

Theoretical work based on the collision history of the solar system suggests an isotropic rotational pole distribution for asteroids (Davis et al., 1989). However, observational data have already suggested otherwise. Analyses based on photometric data have indicated that many rotational poles of small asteroids (D < 30 km) seem to be directed far from the ecliptic plane, and that there exists a preferential prograde rotation, possibly of primordial origin, for large asteroids (D > 60 km) (Kryszczyńska et al., 2007, La Spina et al., 2002, Hanuš et al., 2011). A full explanation of pole depopulation near the ecliptic remains elusive, but the YORP (Yarkovsky-O'Keefe-Radzievskii-Paddack) effect has been indicated as a possible cause.

Other evidence for non-isotropic spin distribution comes from the Koronis family. Based on a 10-asteroid sample and large lightcurve amplitudes, it has been suggested that the Koronis family members have their spin vectors aligned, clustered towards very low or very high obliquities therefore preferentially presenting equatorial aspects to Earth-based observers (Slivan, 2002). The alignment has subsequently been explained (Vokrouhlický et al., 2003) by a combination of the YORP effect and resonances with Saturn. Vokrouhlický et al. (2003) found that prograde rotators should have their spins rates slowed and their spin axis cast into a slow precession and next locked into spin-orbit resonance. Asteroids in this particular equilibrium state are said to be in a so-called Slivan state.

For near-Earth asteroids (NEAs), an excess of retrograde rotators has been reported (La Spina et al., 2004). Retrograde rotators are more likely to be injected into NEA orbits via the so-called Yarkovsky effect. The effect causes the main-belt asteroids (MBAs) in retrograde rotation to drift towards the Sun, injecting them into resonant regions, thence to Earth-approaching orbits.

In general, asteroid spins can be affected by collisional processes, close encounters with planets (Scheeres et al., 2000), tidal effects during close encounters (Richardson et al., 1998), and processes such as the YORP effect (Vokrouhlický and Capek, 2002, Rubincam, 2000, Binzel, 2003). Random collisions among asteroids cause their spin axes to be oriented in random directions (Davis et al., 1989). For small asteroids, the YORP effect is the dominant force influencing asteroid spins. YORP is for example responsible for the spin-up of rubble-pile asteroids, leading to the creation of binary objects (Walsh et al., 2008). YORP also leads to changes in asteroid spin axes.

Much less attention has been paid to the spin-axis longitudes, which are generally thought to follow a uniform distribution (Davis et al., 1989, De Angelis, 1995). It has been argued that precession of orbits has erased any original anisotropy in pole longitudes. De Angelis (1995) has noted that no significant information can be extracted from pole longitudes, because they change due to the precession of rotation axes arising from the tidal torques by the Sun and planets (Burns, 1971). The precession period is orders of magnitude shorter than the age of the solar system (Magnusson, 1986), so no information about the earliest asteroid belt can be obtained from the distribution of pole longitudes. More recent studies also suggest isotropic longitude distributions (for example (Hanuš et al., 2011, Kryszczyńska et al., 2007)).

In this study, we estimate spin-axis longitudes for hundreds of thousands of asteroids using the magnitude method (Magnusson, 1986) and photometric data from the Lowell Observatory database. The method is described in Sec. 2, and the Lowell Observatory database is described in Sec. 3. In Sec. 4, we discuss our results. It should be noted that we present only empirical distributions for the spin-axis longitudes and do not seek to provide possible physical or observational explanations for the shape of the distributions. Conclusions and future research are outlined in Sec. 5.

## 2 Method

We use the so-called magnitude method (Magnusson, 1986) relying on the longitude variation of the mean absolute brightness (an example is given in Fig. 1). In the absence of surface albedo features, it can be assumed that the peak absolute brightness occurs at minimum polar aspect angle; that is, when an asteroid's spin axis is most nearly pointing toward or away from the Earth. We fit a sinusoid to the brightness variations as shown in Fig. 1, and find the spin-axis longitude at the maximum of the curve. The fitted curve is:

$$V(\alpha) = V_0 + \frac{\Delta V}{2} \sin(2\lambda + \lambda_0) \qquad (1)$$

where the phase $\lambda_0$, amplitude $\frac{\Delta V}{2}$, and origin point $V_0$ are fitted simultaneously using least squares and $\lambda$ is the heliocentric ecliptic orbital longitude. This simple spin-axis longitude computation method is well suited to the noisy data at hand. A two-fold ambiguity is present in the method for all objects: there are two equally likely solutions 180 degrees apart in longitude for each object and it is not possible to identify which of the two solutions is true. Therefore the fit assumes that the mean (rotation averaged) brightness of an asteroid with ecliptic longitude is symmetric with 180° and that a symmetric solution in the range 180-360° is also possible. This implies that the distributions of spin-axis longitudes for the 180°- 360° range would look identical to those in the 0°-180° range. To test if the fit is really symmetric with 180° we fit both the first and second harmonics to the data, that is we fit the function described by:

$$V(\alpha) = V_0 + A_1 \sin\lambda + B_1 \cos\lambda + A_2 \sin 2\lambda + B_2 \cos 2\lambda \qquad (2)$$

and check the amplitude of the first harmonic. For 98.3% of the objects, the amplitude is zero within the error bars. For the remaining 1.7% of the objects, the fit including both harmonics over-fits the noisy data. The simple fit (Eq. 1) is therefore a good approximation and can be used in spin-axis longitude computation.

The fit is generally well defined for asteroids exhibiting significant peak-to-peak brightness variation (**>** 0.15 mag), but cannot usually be reliably obtained for asteroids having smaller variation. Asteroids whose poles are perpendicular to the ecliptic exhibit little or no variation; those with poles directed closer to the ecliptic will have larger variation. Most of the fitted objects have their peak-to-peak mean magnitude variation between 0.15 mag and 0.45 mag. In Fig. 4, we plot the distribution of the peak-to-peak magnitude variation. Also, asteroids having small numbers of observations (fewer than 50, say) cannot usually be reliably fitted.

We then estimate the spin-axis longitudes of hundreds of thousands of asteroids, creating the most extensive list of asteroid spin-axis longitudes currently known (the Poznań Observatory database (Kryszczyńska et al., 2007) comprises fewer than two hundred rotational pole solutions).

In Fig. 2, we plot the spin-axis longitudes based on the heliocentric longitude brightness variation (Lowell Observatory database) versus the spin-axis longitudes from the Poznań Observatory database (Kryszczyńska et al., 2007). There is good agreement between the results from the magnitude method and estimates by other authors. Outliers near 0° and 180° arise from asteroids in the Poznań database having spin estimates from authors who do not agree with each other.

Fig. 3 shows the distribution of uncertainty in spin-axis longitude with increasing numbers of observations. Improvement in the longitude fits with increasing numbers of observations is obvious.

The biggest advantage of the magnitude method is its simplicity. The spin-axis longitudes estimates obtained by the method can help constrain the phase-space of possible asteroid spin and shape solutions in more sophisticated methods such as the lightcurve inversion methods, especially in cases where the parameter phase space has many local minima. Kaasalainen et al. (2001) have developed convex inversion methods that, in the case of extensive observational data, converge to a global minimum. For sparse observational data, our spin-axis longitude estimates can help localize the global minimum. The estimates can be especially useful in the analyses of asteroid lightcurves expected from the upcoming large-area sky surveys (e.g., Pan-STARRS, LSST, and the Gaia mission).

## 3 Lowell observatory photometric database

Data from the Lowell Observatory photometric database combines orbital data (from the Lowell Observatory orbital data file maintained by EB and LHW) with photometric data from the Minor Planet Center (MPC). Most of the photometric data are of low precision (generally rounded to 0.1 mag) and low accuracy (rms magnitude uncertainties of 0.2 to 0.3 mag are typical). The MPC data comprise photometric observations from many sources (each having different systematic
and random errors, sometimes time variable). The photometric data are very numerous: in the present study we have used about 47,000,000 individual, largely independent magnitude estimates. For most asteroids, there exist photometric data sampled at a variety of heliocentric longitudes, and therefore different asteroid spin-axis aspects. We have used data from eleven observatories. Most of them have provided photometric data during the course of NEA searches, though the overwhelming majority of the data pertain to MBAs and Jupiter Trojan asteroids. The data were calibrated using accurate broad-band photometry of asteroids observed in the course of the Sloan Digital Sky Survey (SDSS) (Ivezić et al., 2001). The SDSS data were converted to the *V* band using transformations derived by Rodgers et al. (2006). Because of the limited magnitude range of the SDSS data, the brightest (in practice, the first thousand numbered asteroids) and faintest (mostly TNOs, that is, transneptunian objects) are not calibrated. Thus, results for bright and very faint asteroids are less reliable than those whose brightness falls within the range of the SDSS data. A description of the data reduction and calibration can be found in (Oszkiewicz et al., 2011).

## 4 Results

We present histograms of asteroid spin-axis longitudes for different dynamical groups and families. Bins were normalized to sum up to unity. The dynamical classification was extracted from JPL, 2011, and the asteroid family membership was derived from Nesvorný (2010).

| Group | (a) | (b) | (c) | (d) |
|---|---|---|---|---|
| MBA | 17160 | 11650 | 36159 | 109404 |

| Group | (a) | (b) | (c) | (d) |
|---|---|---|---|---|
| NEA | 98 | 31 | 191 | 978 |
| TNO | 1 | 0 | 1 | 9 |
| Mars crossers | 186 | 92 | 369 | 1495 |
| Jupiter trojans | 121 | 71 | 292 | 1102 |
| Centaurs | 5 | 2 | 9 | 30 |

*Table 1: Number of asteroids per dynamical group. a)* $N_{obs} \geq 50, 3 \times \sigma \leq 30º$ *b)* $N_{obs} \geq 100, 3 \times \sigma \leq 30º$ *c)* $N_{obs} \geq 50, 3 \times \sigma \leq 40º$ *d)* $N_{obs} \geq 25, 3 \times \sigma \leq 60º$

Our full sample comprises 355,926 numbered and unnumbered asteroids. The sample reduces to 18,471 asteroids having at least 50 observations and 3 × σ spin-axis longitude uncertainties less than 30º (which corresponds to a moderate 1-σ spin-axis longitude uncertainty of 10º), which we consider a reasonable cut. The sample includes 17,160 MBAs, 98 NEAs, 121 Jupiter Trojans, 186 Mars crossers, 1 TNO, 5 Centaurs, and the members of most asteroid families (please see Table 1 for the number of objects per dynamical group depending on the selection criteria). For TNOs and Centaurs, no statistical study based on such a small number of objects is possible.

There might be some selection biases in our sample, such as those related to the amplitude of brightness variation (only asteroids having their rotational poles close to the ecliptic plane exhibit large enough brightness variation). However, the significance of the biases remains unknown in the present work and cannot be easily estimated.

## 4.1 Dynamical populations

Figure 5 shows the distribution of spin-axis longitudes for MBAs. The longitude distribution for MBAs is far from uniform and shows distinct features: an excess of spin axes in the longitude interval 30º-110º (with two maxima the first one being located between 0º-55º and the second one, more pronounced between 70º-110º) and a paucity between the longitudes 120º-160º. The paucity, the excess, and the second maximum are about 3-σ significance over the level expected from a random distribution and thus can be considered real. The first maximum is only 1-σ above the largest dip between the two intervals and thus cannot be confirmed.

The anisotropy of longitudes for MBAs has already been suggested by La Spina et al. (2003) and Samarasinha and Karr (1998). However that suggestion is contrary to other authors. For example Hanuš et al. (2011) found that the longitude distribution for MBAs shows no significant features and is very close to uniform, with an exception of asteroids smaller than 30 km. Those asteroids have shown a small excess of small spin-axis longitudes, but it was thought to be a random coincidence rather than the result of a physical process. Also, Kryszczyńska et al. (2007) have concluded that the dips in the longitude distributions in the regions 120º -180º and 300º- 360º are only of about 1-σ significance, and thus cannot be confirmed.

Figure 6 shows the distribution of spin-axis longitudes for NEAs. The distribution is clearly different from that for MBAs. In Fig. 6, for NEAs, the distribution exhibits two maxima (the first maximum between 0º-70º and the second between 110º-180º). The two maxima are, however, only about 1-σ above the background, so cannot be confirmed. It has previously been suggested that the NEA longitude distribution exhibits two sharp maxima (Kryszczyńska et al., 2007), but the finding has not been confirmed because of the low contrast of the maxima compared to the mean background.

Asteroids from the Mars orbit crossing population show longitude distribution not far from uniform with only very weak features similar to those of MBAs (see Fig. 7). The features are less pronounced than those of MBAs and only of about 1- σ significance.

Asteroids from the Jupiter Trojan population (Fig. 8) show similar features as the MBAs; that is, two maxima (the first located between 0º- 55º and the second, more pronounced, between 70º- 110º) and a minimum (between 120º- 160º), both of which are also less than 1- σ significance. For the TNO and Centaur population we cannot draw any conclusions due to small-number statistics.

| Group | (a) vs. uniform | (b) vs. uniform | (c) vs. uniform | (d) vs. uniform |
|---|---|---|---|---|
| MBA | 0.1, $10^{-94}$ | 0.1, $10^{-82}$ | 0.1, $4\times10^{-131}$ | 0.1, $4\times10^{-160}$ |
| NEA | 0.13, 0.42 | 0.14, 0.7 | 0.08, 0.75 | 0.08, 0.6 |
| Mars crossers | 0.12, 0.12 | 0.1, 0.6 | 0.1, 0.08 | 0.1, 0.14 |
| Jupiter trojans | 0.2, 0.002 | 0.2, 0.11 | 0.2, $0.8\times10^{-4}$ | 0.2, 0.0008 |
| Centaurs | 0.8, 0.4 | 0.8, 0.4 | 0.5, 0.4 | 0.5, 0.1 |

*Table 2: Statistical p-values from the K-S test for the different samples, (a), (b),(c), (d) as in Table 1.*

To test the robustness of all the distributions, we plot the longitude distributions based on different cut-offs for the numbers of observations and the 3 × σ longitude uncertainty. Next, we use the Kolmogorov-Smirnov (K-S) test to examine the randomness of all the distributions. The null hypothesis that the distribution being tested is uniform is rejected or accepted based on K-S statistics and p-values. The obtained K-S statistics and p-values are listed in Table 2. If the K-S statistic is small or the p-value is high (> 0.05), then we cannot reject the hypothesis that the two distributions are the same. The null hypothesis can be clearly rejected for MBAs. For MBAs, both the K-S statistics are large and the p-value is small. The spin-axis longitude distribution for MBAs is therefore nonrandom. For NEAs and Mars crossers, the null hypothesis cannot be rejected. For Jupiter Trojans, the null hypothesis can be rejected for cases (a), (c), and (d). Condition (b) is the most strict one, and therefore we are inclined to conclude that the spin-axis longitude distribution for the Jupiter Trojans is non-random.

A clear explanation for the shape of the longitude distributions is missing. However it is worth mentioning that the distribution for the MBAs is different that that for NEAs. Although it can not be statistically shown, the remaining groups (Jupiter trojans, Mars crossers) show similar trends in their longitude distributions. Therefore a possible mechanism has to explain the lack of (or smaller) influence on the NEAs. To test if YORP could influence the distributions, we plotted the spin-axis longitudes for different absolute magnitude regimes: First, corresponding to an interval of H = (0 – 9.5) mag, that is asteroids larger than approximately 30 km; Second corresponding to an interval  H = (9.5 - 11.5) mag for asteroids with moderate sizes between 30 km and 10 km; and the third one H>11.5 mag corresponding to asteroids smaller than 10 km. The trend in the spin-axis longitude distribution is visible in all the regimes. YORP is therefore unlikely to be the main explanation for the shape of the observed longitude distribution.

## 4.2 Asteroid families

We plotted the spin-axis longitudes for all the asteroid families contained in our data set. Here we present the distributions for a few selected families: (158) Koronis - Fig. 9, (4) Vesta - Fig. 10, (8) Flora - Fig. 11, and (44-142) Nysa-Polana - Fig. 12.

The longitude distribution for the Koronis family is unimodal, with an excess of longitudes between 60º-110º, which could be a reflection of the general trend visible for MBAs.

Eight of the Slivan's Koronis asteroids can be compared with our results. The computed spin-axis longitudes are: (2953) Vysheslavia: 103º +/- 21.7º, (1223) Neckar: 69.3º +/- 8º, (720) Bohlinia: 48.4º +/- 14.5º, (534) Nassova: 61.9º +/- 14.7º, (321) Florentina: 91.6º +/- 19.7º, (311) Claudia: 66.8º +/- 26.4º, Dresda: 101.9º +/-18.2º and (243) Ida: 72.7º +/- 24º. All of those agree with previously estimated values (Slivan et al. 2002, Slivan et al. 2009) except for (311) Claudia and (2953) Vysheslavia (Slivan et al. 2009 , Vokrouhlický et al. 2006). For those objects, the literature values of the spin-axis longitude are 24º +/-5º and 11º +/-10º respectively. Both of those objects have large numbers of observations made at various observatories. (311) Claudia has 442 observations taken at 11 different observatories (observatory codes: 689, 699, 608, 704, 644, 703, 333, 691, 1412, 683, 1696). (2953) Vysheslavia has 658 observations made at 9 different observatories (observatory codes: 691, 699, 704, 703, 608, 1696, 645, 1412, 644). Both of the objects also have the mean peak-to-peak magnitude variation above 0.15 mag. Therefore both of the objects were considered as an acceptable fit, however, visual inspection of the fit shows large scatter of the data. We consider the number of this kind of invalid fits to be small and not affecting the overall distributions. However in future some automatic testing rather than visual inspection could possibly be developed to detect those sort of cases among the large amount of data at hand.

Spin distributions for families Vesta, Flora, and Nysa-Polana seem to follow the general trend of MBAs, that is a paucity between 120º-160º and an excess of longitudes between 70º-110º. Most of the families for which we have at least 100 spin longitudes also follow the general shape of the MBA distribution.

# 5 Conclusions and future work

We have estimated spin-axis longitudes for hundreds of thousands of asteroids, based on the brightness variation with ecliptic longitude and using the Lowell Observatory photometric database. The number of spin-axis longitudes computed is an enormous increase in the number of previously known asteroid spin-axis longitudes. The estimated spin-axis longitudes are publicly available online on Planetary System Research group – University of Helsinki webpages (https://wiki.helsinki.fi/display/PSR/Planetary+System+Research+group) and on an ftp site at Lowell Observatory ftp://ftp.lowell.edu/pub/elgb/summary.out. Based on the spin-axis longitude distributions for MBAs, we concluded that the distribution is far from uniform, with an excess at longitudes 30°-110° and a paucity between longitudes 120°-160°. Longitude anisotropy is consistent with La Spina et al. (2003), Samarasinha and Karr (1998) and contradictory to Kryszczyńska et al. (2007), Hanuš et al. (2011). Anisotropy of the longitude distributions was not confirmed in other dynamical groups, except for Jupiter Trojans, which exhibit features similar to MBAs. We also investigated asteroid families. For the Koronis family, we showed that spin-axis longitudes are clustered around 60°-110°. Spin-axis distributions for most other asteroid families reflect the features visible in the MBAs distribution. Explanation of the physical causes for the shape of the distributions is beyond of the scope of this paper, and will require extensive modeling of the YORP effect, precession and observational selection effects.

# 6 Acknowledgments

Research has been supported by the Magnus Ehrnrooth Foundation, Academy of Finland (contract 127461), Lowell Observatory, Polish National Science Center (grant number 2012/04/S/ST9/00022), and the Spitzer Science Center. We would like to thank David Vokrouhlický, David Nesvorný, and Alan Harris (Space Science Institute) for valuable comments and discussion. We would also like to thank our reviewers Petr Pravec and Agnieszka Kryszczyńska for insightful and detailed comments.

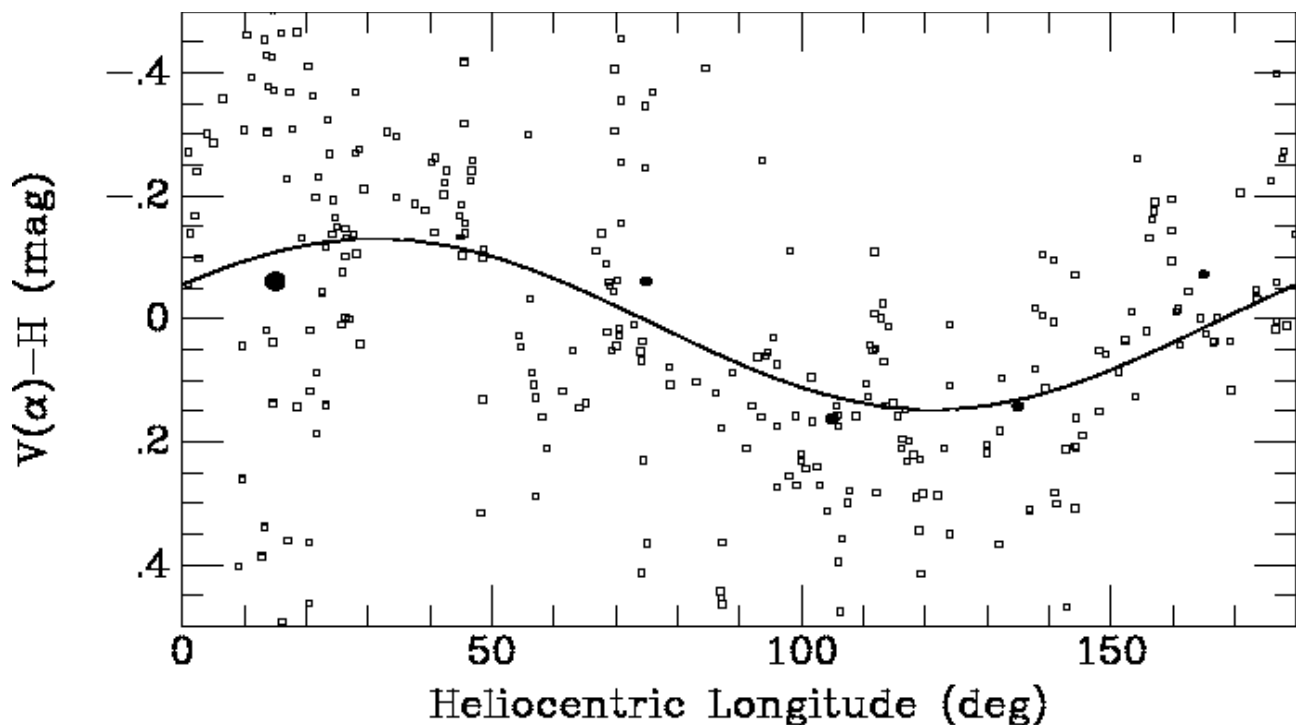

*Illustration 1: Brightness variation with heliocentric longitude for (93) Minerva. Filled circles depict brightness variation averaged over 30° intervals; symbol size is proportional to the number of observations. The least-squares fit indicates a spin-axis longitude of 33.3° ±13.3° (or a symmetric solution of 213.3°).*

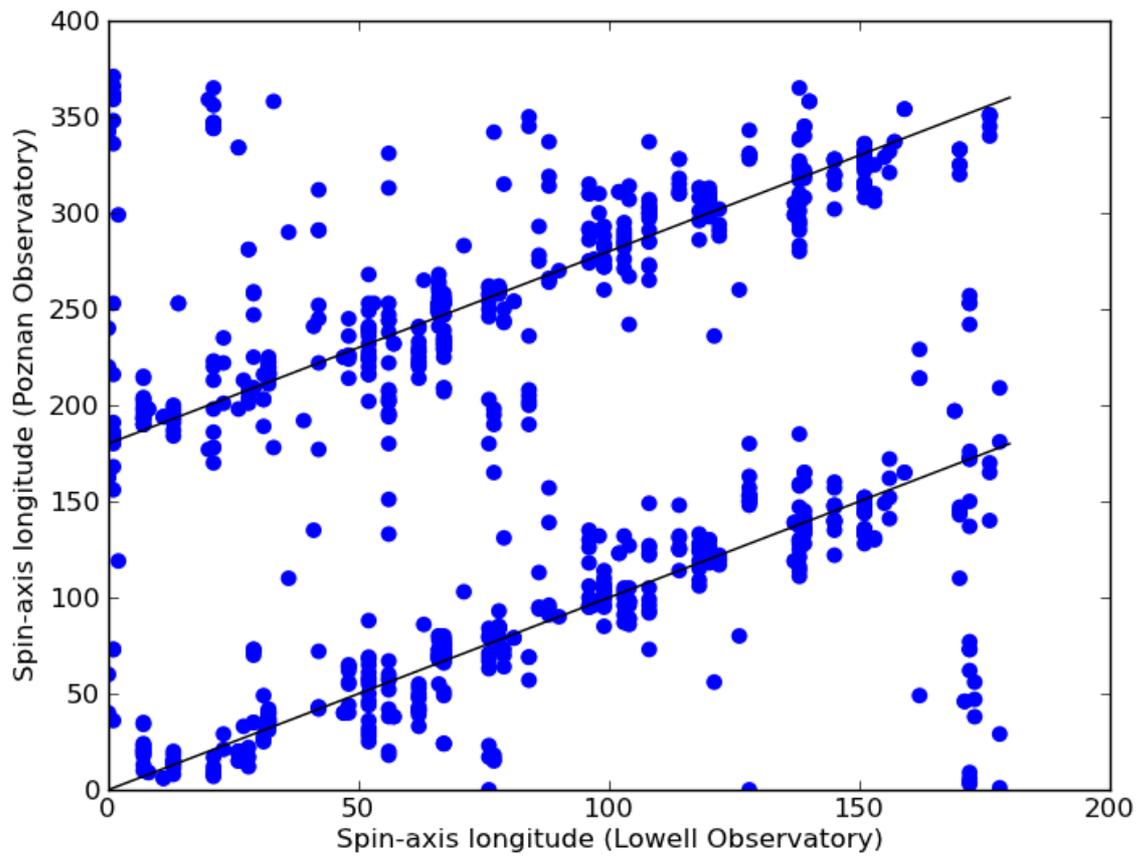

*Illustration 2: Comparison of estimated spin-axis longitudes derived from brightness variation as a function of orbital longitude (Lowell Observatory) and by using other methods (Poznań Observatory).*

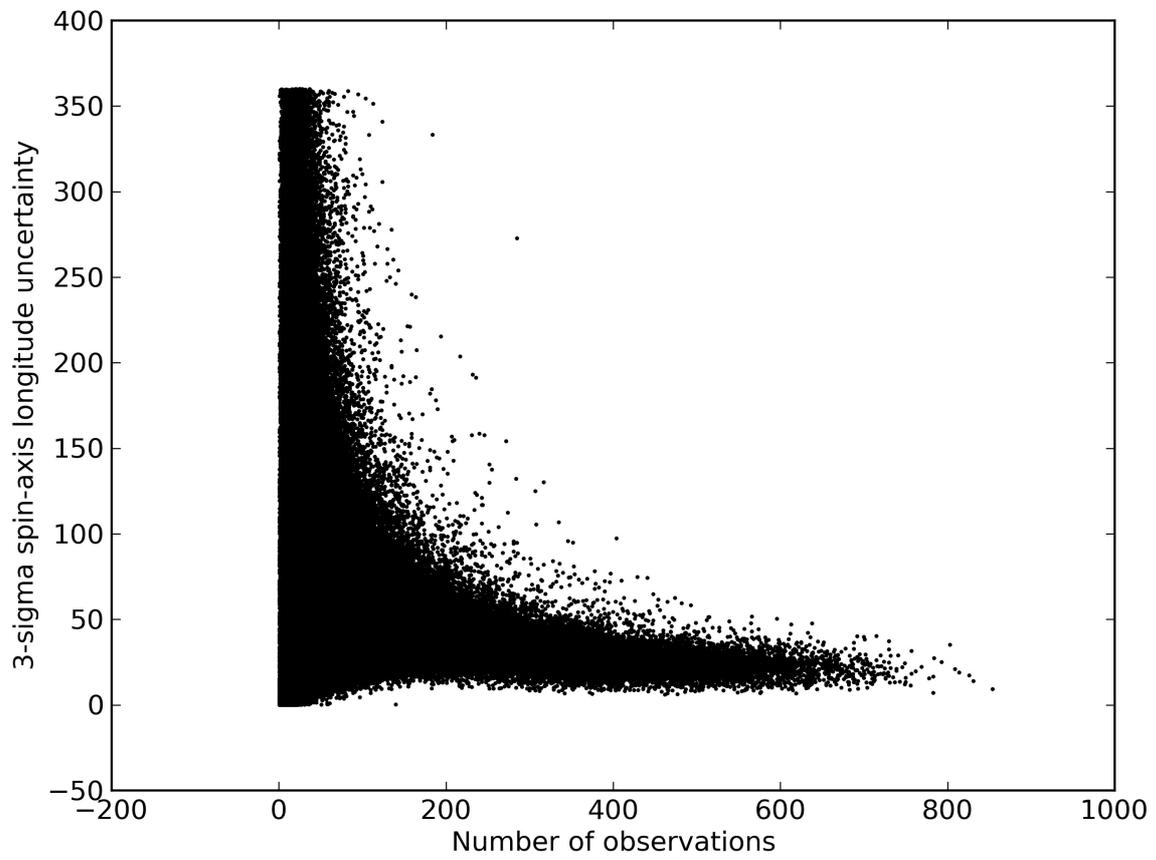

*Illustration 3: 3xσ spin-axis longitude uncertainty versus the number of observations.*

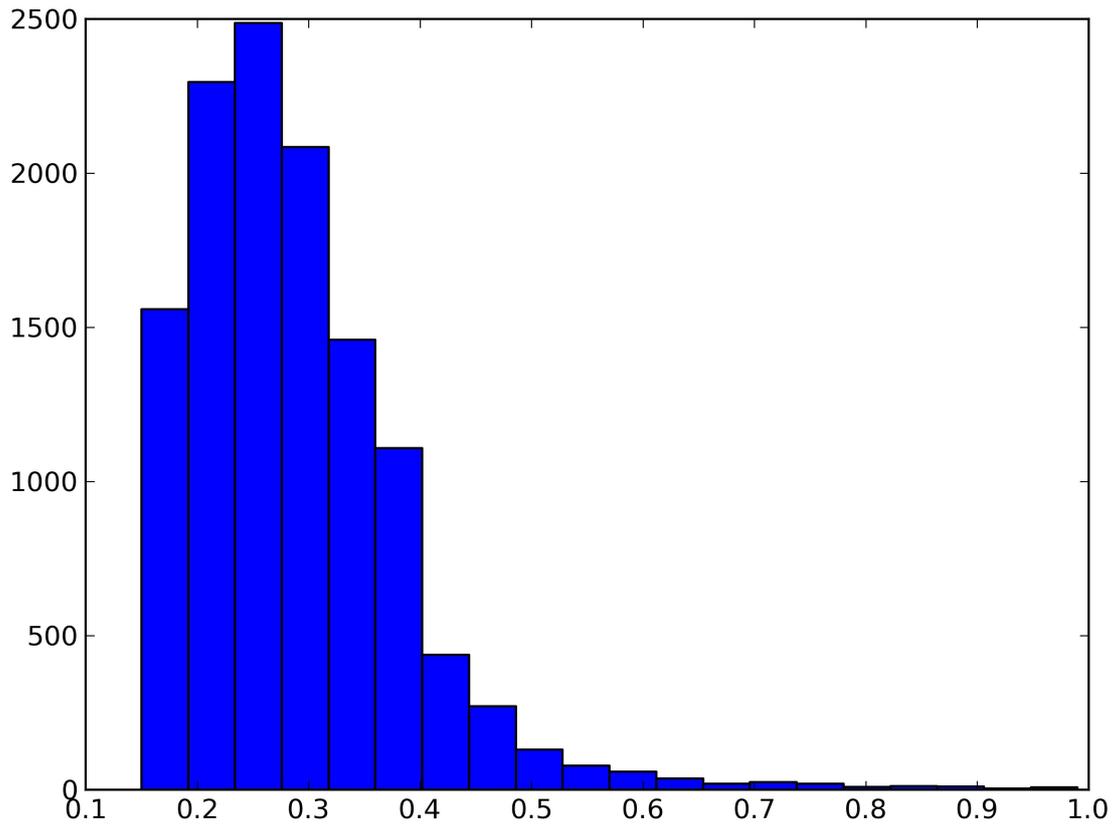

*Illustration 4: Distribution of the peak-to-peak magnitude fit values*

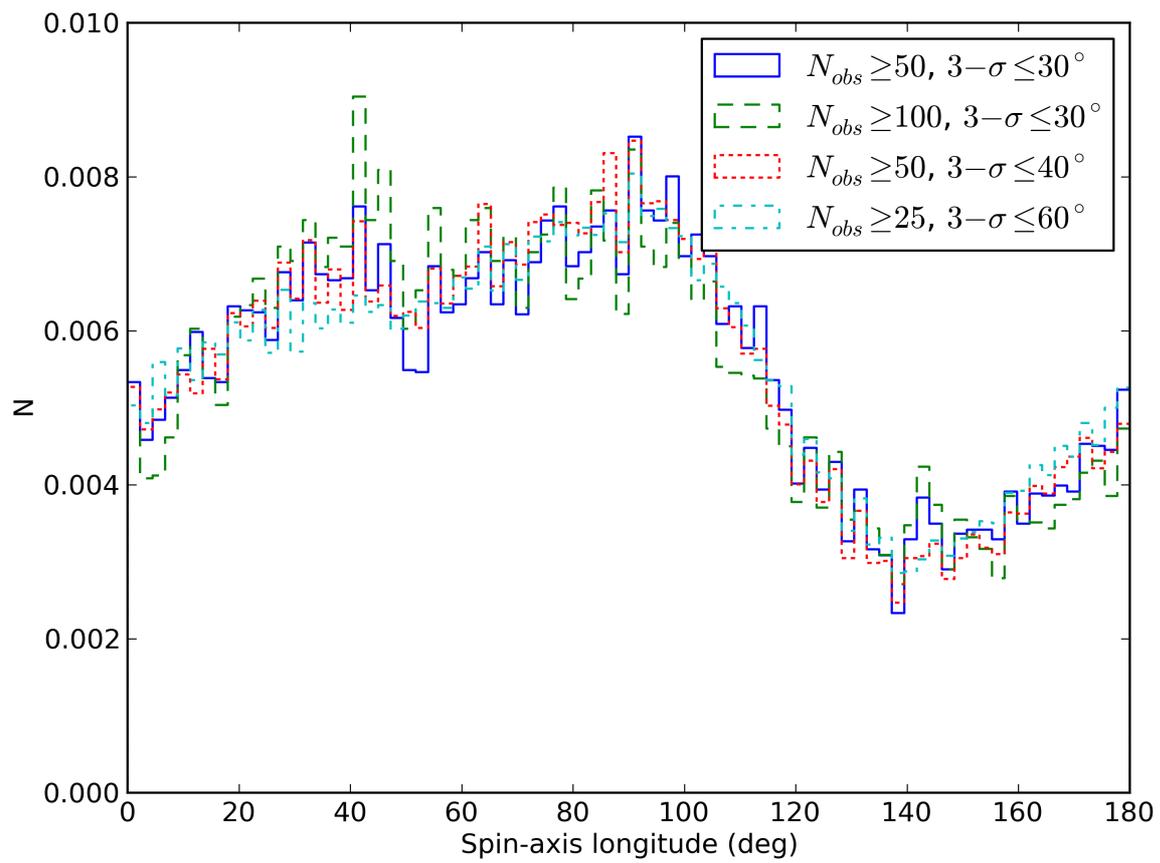

*Illustration 5: Spin-axis longitude distribution for main-belt asteroids.*

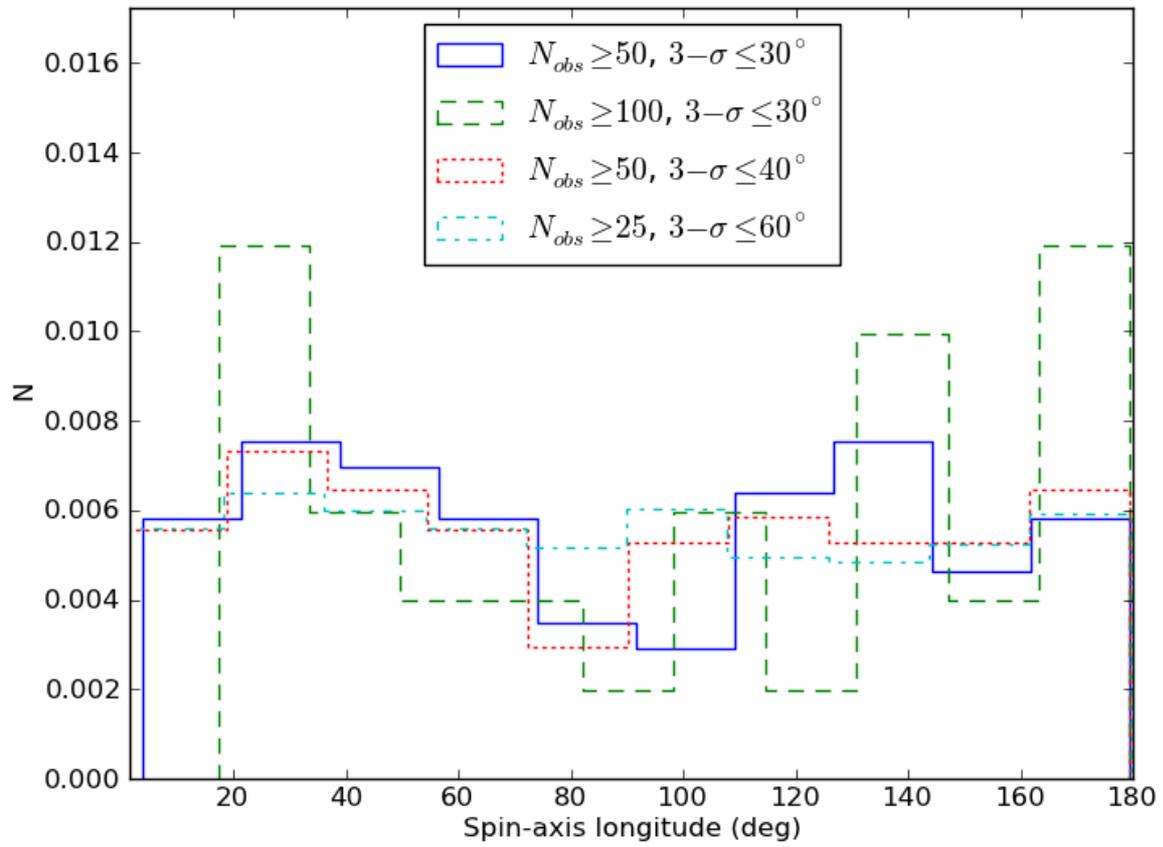

Illustration 6: Spin-axis longitude distribution for near-Earth asteroids.

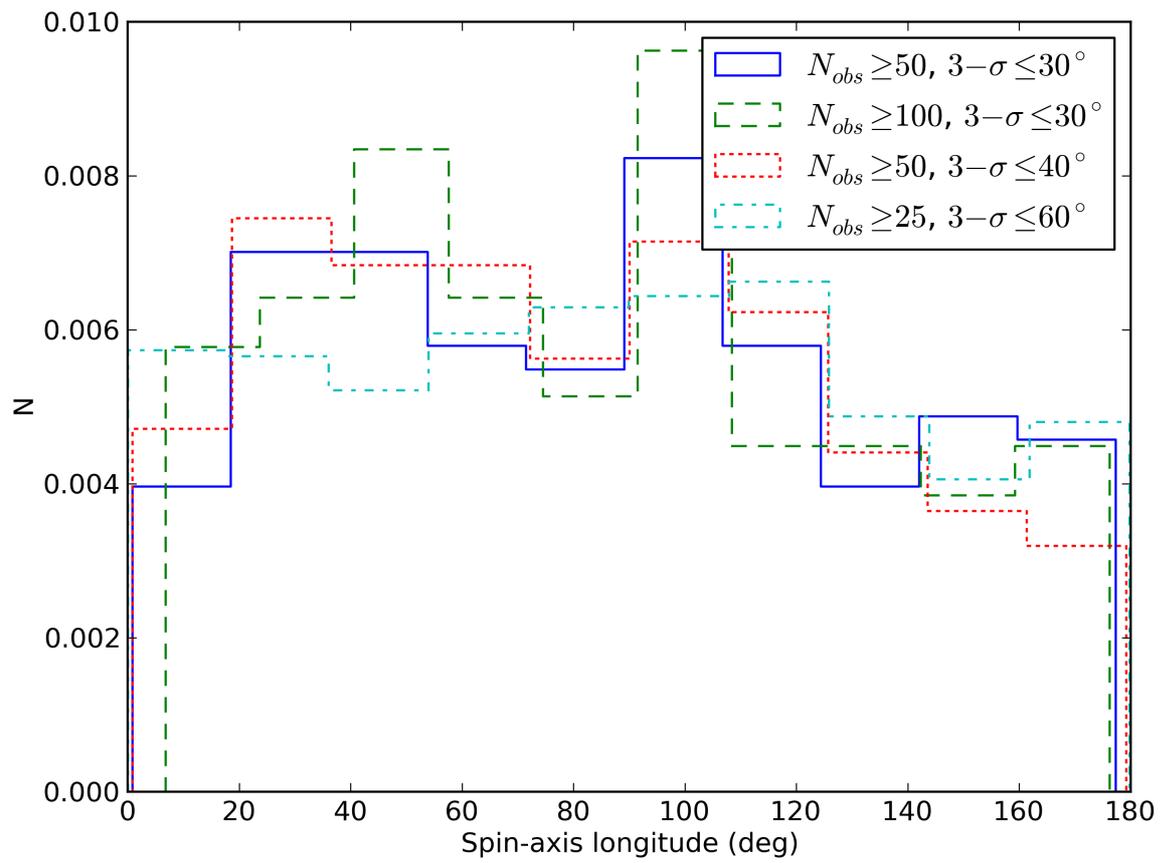

*Illustration 7: Spin-axis longitude distribution for Mars crossers.*

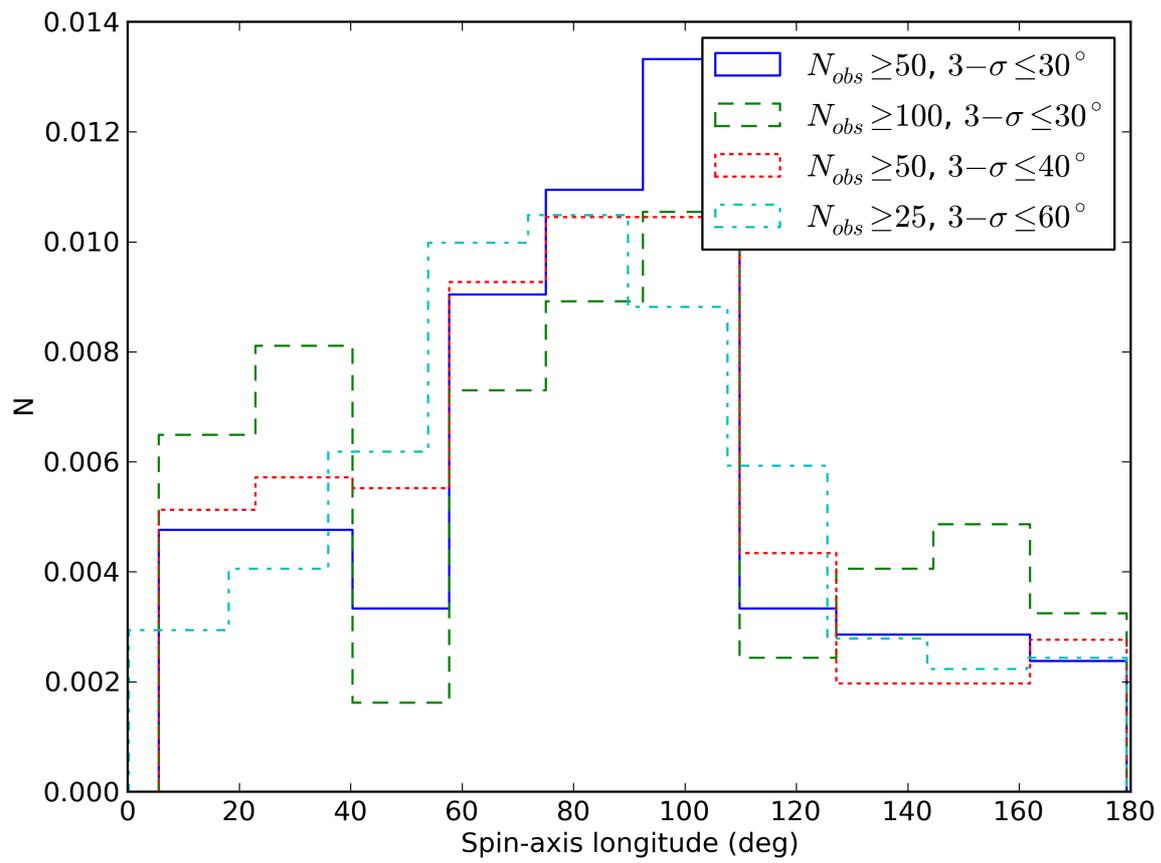

*Illustration 8: Spin-axis longitude distribution for Jupiter Trojans.*

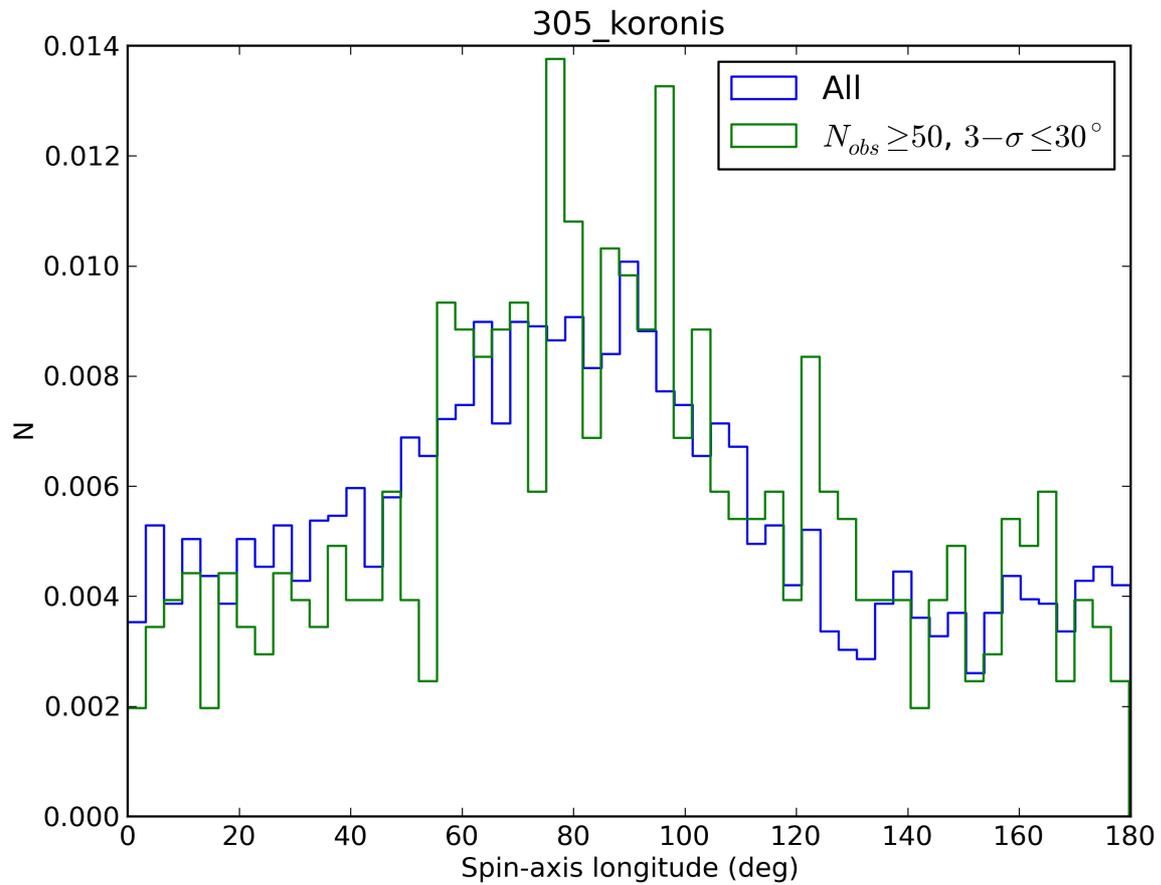

*Illustration 9: Spin-axis longitude distribution for the Koronis family. The figure title corresponds to the name of the file containing family members in the HCM asteroid families v2.0 database Nesvorný, (2010).*

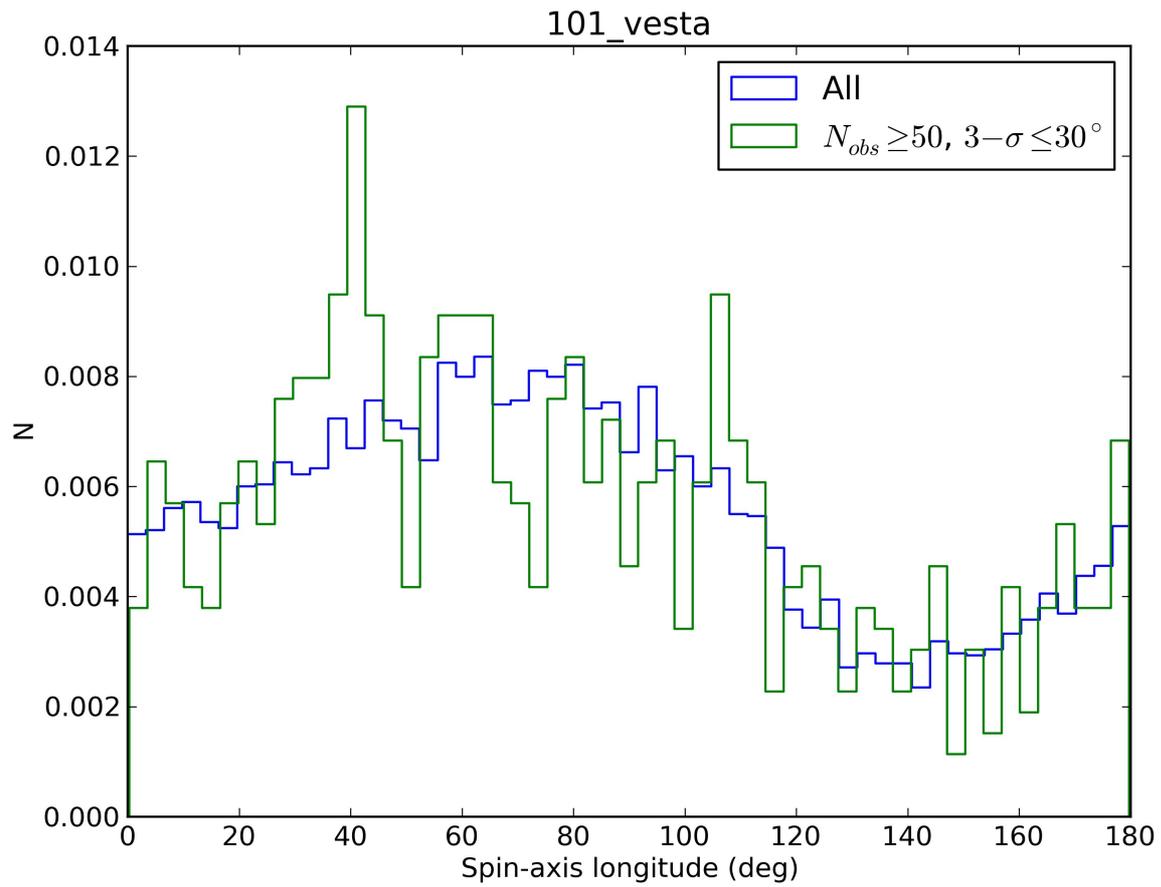

*Illustration 10: Spin-axis longitude distribution for the Vesta family. The figure title corresponds to the name of the file containing family members in the HCM asteroid families v2.0 database Nesvorný, (2010).*

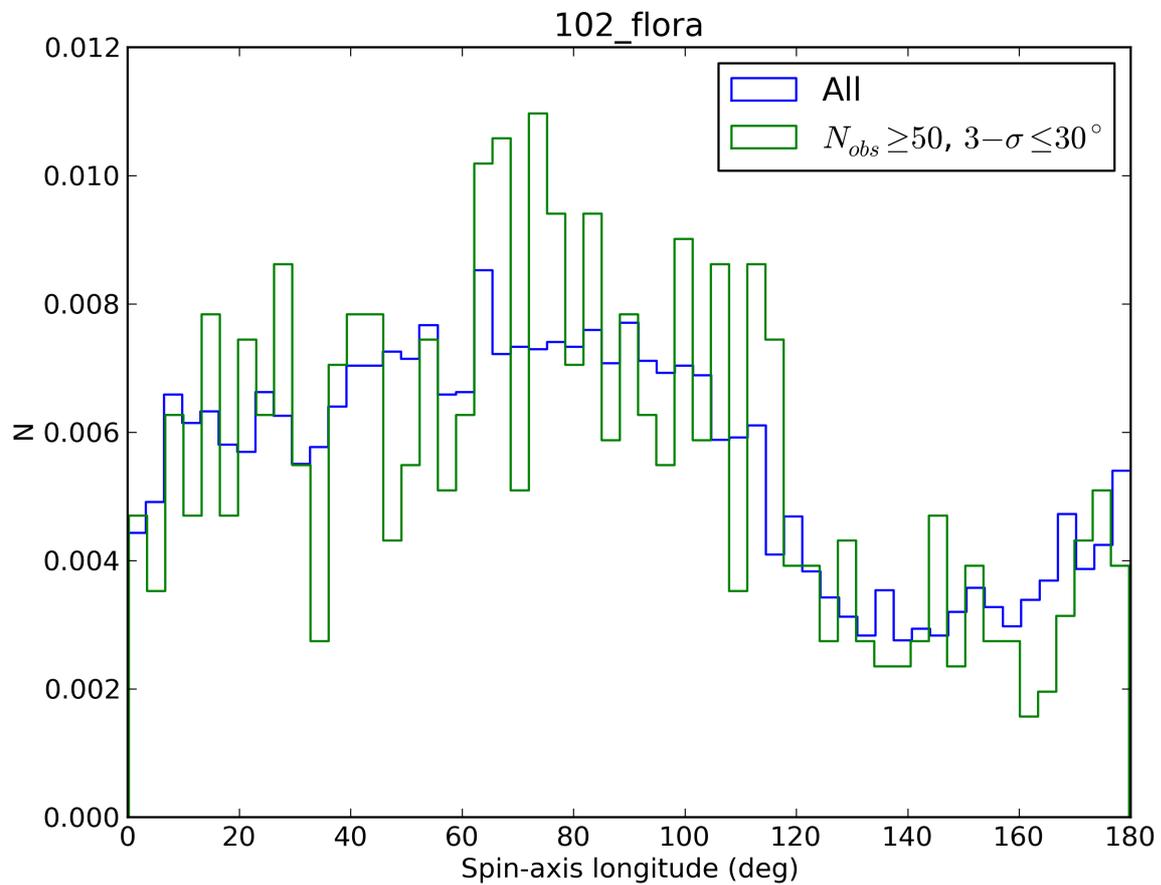

Illustration 11: Spin-axis longitude distribution for the Flora family. The figure title corresponds to the name of the file containing family members in the HCM asteroid families v2.0 database Nesvorný, (2010).

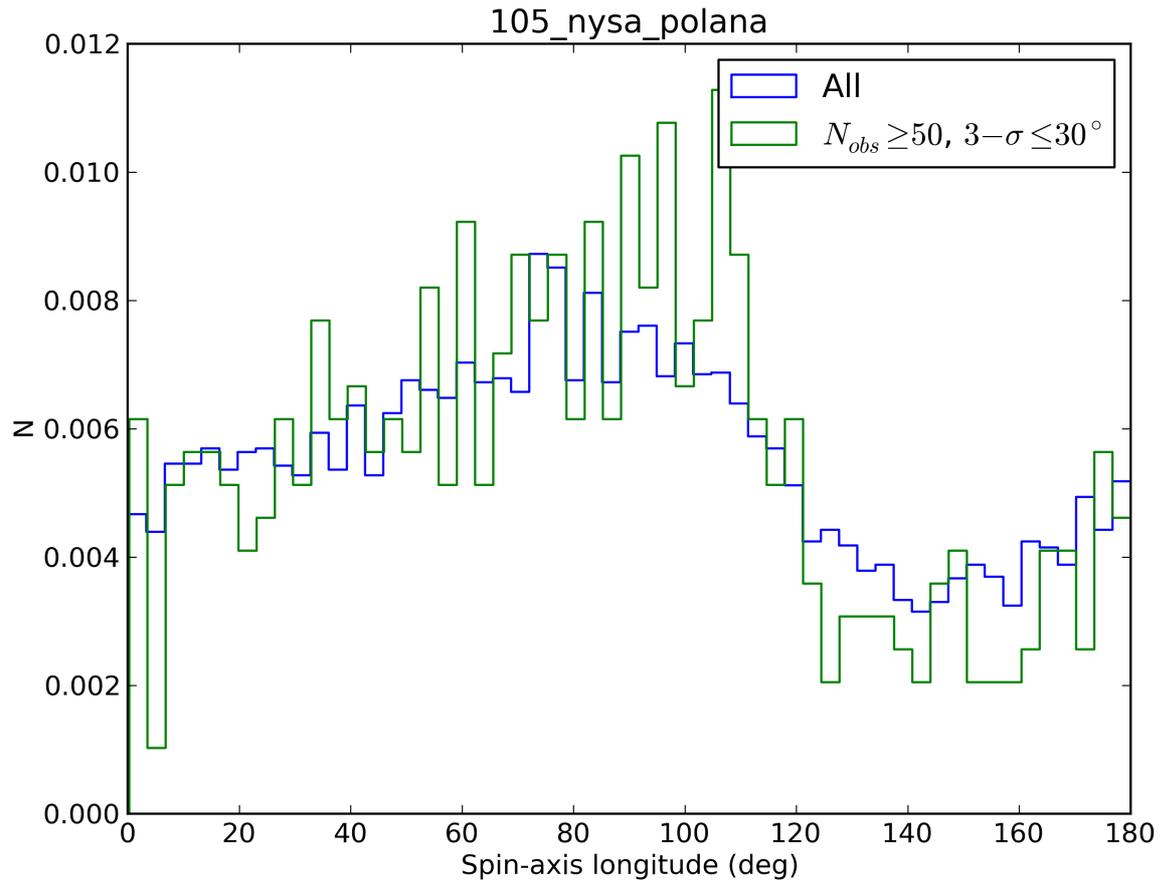

*Illustration 12: Spin-axis longitude distribution for the Nysa Polana family. The figure title corresponds to the name of the file containing family members in the HCM asteroid families v2.0 database Nesvorný, (2010).*